\documentclass[12pt,preprint]{aastex}

\slugcomment{}

\shorttitle{COMs in G34.43+00.24 MM3}
\shortauthors{Sakai et al.}

\begin{document}


\title{ALMA Observations of the IRDC Clump G34.43+00.24 MM3: Complex Organic and Deuterated Molecules}


\author{Takeshi Sakai\altaffilmark{1}, Takahiro Yanagida\altaffilmark{1}, Kenji Furuya\altaffilmark{2}, Yuri Aikawa\altaffilmark{3}, Patricio Sanhueza\altaffilmark{4}, Nami Sakai\altaffilmark{5}, Tomoya Hirota\altaffilmark{4}, James M. Jackson\altaffilmark{6}, and Satoshi Yamamoto\altaffilmark{7}}

\altaffiltext{1}{Graduate School of Informatics and Engineering, The University of 
Electro-Communications, Chofu, Tokyo 182-8585, Japan.}
\altaffiltext{2}{Center for Computer Sciences, University of Tsukuba, Tsukuba 305-8577, Japan}
\altaffiltext{3}{Department of Astronomy, The University of Tokyo, Bunkyo-ku, Tokyo 113-0033, Japan}
\altaffiltext{4}{National Astronomical Observatory of Japan, National Institutes of Natural Sciences, 2-21-1 Osawa, Mitaka, Tokyo 181-8588, Japan}
\altaffiltext{5}{The Institute of Physical and Chemical Research (RIKEN), 2-1, Hirosawa, Wako-shi, Saitama 351-0198, Japan}
\altaffiltext{6}{School of Mathematical and Physical Sciences, University of Newcastle, University Drive, Callaghan NSW 2308, Australia}
\altaffiltext{7}{Department of Physics, The University of Tokyo, Bunkyo-ku, Tokyo 113-0033, Japan}


\begin{abstract}
We have observed complex organic molecules (COMs) and deuterated species toward a hot core/corino (HC) associated with the infrared dark cloud (IRDC) clump G34.43+00.24 MM3 with the Atacama Large Millimeter/submillimeter Array (ALMA).
We have detected six normal-COMs (CH$_3$OH, CH$_3$CHO, CH$_3$CH$_2$CN, CH$_3$OCH$_3$, HCOOCH$_3$, and NH$_2$CHO), one deuterated-COM (CH$_2$DCN), and two deuterated fundamental molecules (D$_2$CO and DNC) toward G34.43+00.24 MM3 HC. None of these lines, except for CH$_3$OH, are detected toward the shocked regions in our data, which suggests that COMs do not originate in shocks.
The abundance of the COMs relative to CH$_3$OH in G34.43+00.24 MM3 HC is found to be similar to those in high-mass hot cores, rather than those in hot corinos in low-mass star-forming regions.
This result suggests that the physical conditions of the warm-up phase of G34.43+00.24 MM3 HC are similar to those of high-mass sources. 
On the other hand, the D$_2$CO abundance relative to CH$_3$OH in G34.43+00.24 MM3 HC is higher than that of other hot cores, and seems to be comparable to that of hot corinos.
The relatively high D$_2$CO/CH$_3$OH ratio of G34.43+00.24 MM3 HC implies a long cold starless phase of G34.43+00.24 MM3 HC.
\end{abstract}





\section{Introduction}

G34.43+00.24 is a filamentary infrared dark cloud (IRDC) (Figure 1a; Garay et al. 2004; Rathborne et al. 2005; 2006; Sanhueza et al. 2010), and MM3 is the third most massive clump in the IRDC. The mass of MM3 is reported to be 300 $M_\odot$ in the distance of 3.7 kpc (Rathborne et al. 2006; see below for the distance).
Toward MM3, we observed several molecular lines and continuum emission by using the Atacama Large Millimeter/submillimeter Array (ALMA) (Sakai et al. 2013, 2015; Yanagida et al. 2014). We found several continuum sources in this clump, and detected compact emission of a high-excitation methanol line ($E_u$=165 K) toward the most luminous continuum source in MM3 (Figure 1b); this source corresponds to ``Peak A'' in Sakai et al. (2013).  This methanol emission is likely to trace a hot core/corino around a protostar embedded in the continuum source.
In addition, we found a collimated and compact outflow associated with the continuum source (Figure 1b).
Based on the CS and N$_2$H$^+$ emission, we found evidence that the compact outflow is interacting with the ambient dense gas; the CS emission tracing the outflow is anti-correlated with the N$_2$H$^+$ emission tracing the quiescent dense gas (Yanagida et al. 2014; Sakai et al. 2015).
Furthermore, the SiO emission (Sakai et al. 2013) and class I CH$_3$OH masers (Yanagida et al. 2014), which are shock tracers, are detected toward the interacting regions.
Thus, the outflow is likely very young.
In this paper, we investigate the nature of the hot core/corino in G34.43+00.24 MM3 from its chemical composition.

The observationally derived physical parameters of the hot core/corino, such as mass, size, and the dynamical timescale of the outflow, depend on the assumed distance to the source.
Kurayama et al. (2011) derived the distance to be 1.56 kpc from VLBI parallax observations. On the other hand, the distance is estimated to be 3.5-3.9 kpc by assuming a Galactic rotation curve (kinematic distance: Sakai et al. 2008; Sanhueza et al. 2012) and by using the near infrared extinction method (Foster et al. 2012). Foster et al. (2012, 2014) questioned the parallax distance derived by Kurayama et al. (2011), and Xu et al. (2016) suggested that the IRDC G34.43+00.24 is a part of the large molecular complex, including the G34.26+0.15 UC HII region, whose distance is 3.7 kpc (Kuchar \& Bania 1994). 
Thus, the distance to this object is not well defined.
In this paper, we consider both possibilities.

If the distance is 1.56 kpc (the near case),  the gas mass derived with ALMA is less than 1.1 $M_\odot$ (the upper limit value is due to the assumption of the temperature) and the size of the high excitation methanol emission is 800 au $\times$ 300 au. 
The dynamical timescale of the outflow is less than 740 yr (the upper limit value is due to the assumption of the inclination angle; Sakai et al. 2013).
Thus, in spite of the youth of the outflow, the size of the hot core/corino is larger than that of the typical hot corinos in low-mass star forming regions ($<$100 au; Maret et al. 2004).

If the distance is 3.9 kpc (the far case), the gas mass is less than 6.8 $M_\odot$ and the size of the high excitation methanol emission is 1900 au $\times$ 700 au. The dynamical timescale of the outflow is less than 1900 yr.
In this case, the size of the hot core is larger than other intermediate-mass hot cores (300-1200 au; Palau et al. 2011).

Therefore, the size of the hot core/corino in G34.43+00.24 MM3 seems higher than those of the hot corinos and/or intermediate-mass hot cores, in spite of its young age inferred from the outflow. In this paper, we characterize the chemical composition of the hot core/corino in G34.43+00.24 MM3 by using the ALMA data in detail, and compare the results with those of high-mass hot cores and low-mass hot corinos to explore the nature of the hot core/corino.
In particular, we focus on the abundance of N-bearing COM, CH$_3$CH$_2$CN, which could be more abundant in high-mass sources than in low-mass sources (e.g. Taquet et al. 2015).
We also investigate the formation processes of the observed complex organic molecules (COMs) and deuterated molecules.
Hereafter, we refer to the hot core/corino in G34.43+00.24 MM3 as ``G34.43 MM3 HC''.

\section{OBSERVATIONS}

We observed G34.43+00.24 MM3 with ALMA Band 6 and Band 7 in 2012.
The phase center was (R.A.(J2000), Dec.(J2000)) = (18$^{\rm h}$53$^{\rm m}$20.4$^{\rm s}$, 1$^{\circ}$28$^{\prime}$23.0$^{\prime\prime}$).
The data used in this paper are a part of the same observations presented in Sakai et al. (2013), Yanagida et al. (2014), and Sakai et al. (2015). 
The details of the observations are also described in these papers.

In the present paper, we use data from nine spectral windows; Windows 1-7 and Windows 8-9 correspond to ALMA Band 6 and Band 7, respectively. 
A part of the data in Window 2-6 (CH$_3$OH, CS, $^{13}$CS, HCOOCH$_3$, CH$_3$CH$_2$CN, and SiO lines), Window 1 and 7 (DNC and HN$^{13}$C lines), and Window 9 (CH$_3$OH $J_K$=9$_{-1}$-8$_{0}$ $E$ line) are previously reported in Sakai et al. (2013), Sakai et al. (2015), and Yanagida et al. (2014), respectively.
Although the spectrometers were used with the 234 MHz mode with a 61 kHz channel width, we averaged channels in order to reduce the noise level; 40 channels (channel width of 2.44 MHz, corresponding to 3.2 km s$^{-1}$ at 230 GHz) for the data presented in Figure 2, and 24 channels (channel width of 1.464 MHz, corresponding to 1.92 km s$^{-1}$ at 230 GHz) for the data presented in Figure 3.
The observed frequencies, the synthesized beam sizes and the rms noise level are listed in Table 1. 
The data were reduced by the CASA software package. Self-calibration is not applied.

\section{RESULTS \& ANALYSIS}

\subsection{Spectra \& Integrated Intensity Maps} 

Figure 2 shows the spectra toward the peak position of the CH$_3$OH $J_K$=10$_2$--9$_3$ $A^-$ emission in G34.43+00.24 MM3 at (R.A., Dec.) = (18$^{\rm h}$53$^{\rm m}$20.62$^{\rm s}$, 1$^{\circ}$28$^{\prime}$25$^{\prime\prime}$.5), which coincides with the peak position of the dust continuum emission at the present resolution ($\sim$0$^{\prime\prime}$.8$\times$0$^{\prime\prime}$.6). 
We detected 16 species, including six normal-COMs (CH$_3$OH, CH$_3$CHO, CH$_3$CH$_2$CN, CH$_3$OCH$_3$, HCOOCH$_3$, and NH$_2$CHO), one deuterated-COM (CH$_2$DCN), and two deuterated fundamental molecules (D$_2$CO and DNC), above the 3 sigma root-mean-square (rms) noise level. The detected COM lines, except for CH$_2$DCN, are often observed toward high-mass hot cores. COMs are considered to form efficiently on grain surfaces at temperatures ($T$) larger than a few times 10 K, and sublimate to the gas phase at $T$ $\gtrsim$ 100 K, although the actual sublimation temperature would slightly vary among molecules (Garrod \& Herbst 2006; Herbst \& van Dishoeck 2009). Deuterated species are also often detected in hot cores/corinos, which verifies the scenario that COMs are formed at relatively low temperature and/or their parent molecules originate in cold prestellar stage (e.g. Rodgers \& Millar 1996; Aikawa et al. 2012). The highly excited lines and deuterated COMs detected in MM3 are consistent with the properties of hot cores.
The upper state energies of the detected molecular lines range from 22 K to 338 K, where the highest one is for the CH$_3$OH $J_K$=16$_2$--15$_3$ $E$ line. The detections of the high-excitation lines indicate that there is a hot region.
The detected molecular lines are listed in Table 2.

Figures 1c-j show integrated intensity maps of the COMs and deuterated molecules around G34.43 MM3 HC. 
For the NH$_2$CHO, DNC, and D$_2$CO lines, the velocity range for the integration is $\pm$5 km s$^{-1}$ centered at the LSR velocity of 59.7 km s$^{-1}$. For the CH$_3$CH$_2$CN, CH$_3$CHO, HCOOCH$_3$, CH$_3$OCH$_3$, and CH$_2$DCN lines, several transitions are overlapped and unresolved (Table 2).
In such cases, the velocity range for the integration is set from -5 km s$^{-1}$ of the lowest velocity line to +5 km s$^{-1}$ of the highest velocity line. 
In Figure 1, emission of all the COMs is found to be peaked toward G34.43 MM3 HC. 
Towards the interacting region between the outflow and dense gas (CS peak), on the other hand, SiO and class I CH$_3$OH masers are detected, which suggests the existence of shock there (Sakai et al. 2013; Yanagida et al. 2014).

We also found no clear difference in the distribution between O-bearing COMs and N-bearing COMs at the present resolution. This is different from the situation in the Orion KL region; the N-bearing COMs, such as CH$_3$CH$_2$CN, tend to peak toward ``hot core'', while the O-bearing COMs, such as CH$_3$OCH$_3$ and HCOOCH$_3$, rather tend to peak in the so-called ``compact ridge'' (e.g., Blake et al. 1996). 
Although the origin of the difference is still under debate, Blake et al. attributed this spatial distribution to chemical reactions; molecules formed by neutral-neutral reactions dominate in the hot core, while molecules formed by ion-neutral reactions are found in the compact ridge.
Note that, even if the Orion KL region were located at a distance of 3.9 kpc, the hot core and the compact ridge would be resolved with the 0$^{\prime\prime}$.8 beam of our current observations.
Thus, the difference between G34.43 MM3 HC and Orion-KL is not due to the spatial resolution. We will discuss the origin of the N-bearing and O-bearing molecules in G34.43 MM3 HC in Section 4.1.2.

We fit a 2D Gaussian to the integrated intensity maps of the COMs emission near G34.43 MM3 HC in Figure 1, using the Gaussian fitting task in $imview$ of the CASA software package.
The derived source size of the COMs emission is comparable to or slightly larger than the beam size (Table 3).
When we deconvolve the emission with the synthesized beam by using the Gaussian fitting task in $imview$, the size of the CH$_3$CH$_2$CN and HCOOCH$_3$ emission is comparable to that of CH$_3$OH, and the NH$_2$CHO and CH$_3$OCH$_3$ emission is recognized as a point source. On the other hand, the emitting region of the CH$_2$DCN line looks more extended than that of CH$_3$OH.
Although the distribution of CH$_2$DCN seems to be slightly different among the observed COMs, we need a higher angular resolution in order to investigate the spatial differences in the emitting regions in more detail.
As for the other deuterated spices, we found that the distribution of DNC is different from those of D$_2$CO and CH$_2$DCN.  The D$_2$CO and CH$_2$DCN emission peaks toward G34.43 MM3 HC, while the DNC peak is offset from G34.43 MM3 HC (this offset has been reported by Sakai et al. (2015)).

\subsection{Velocity Structures}

Figure 3a shows a moment 1 map of the CH$_3$OH $J_K$=10$_2$--9$_3$ $A^-$ line with the velocity range of 50-70 km s$^{-1}$, and it clearly reveals a significant velocity gradient from west to east.
This gradient is almost perpendicular to the axis of the collimated outflow, suggesting that the CH$_3$OH emission could trace a rotating structure around the protostar.

In Figure 3b, we show the position-velocity diagram through the solid line indicated in Figure 3a. Although we cannot see a clear rotational feature in this figure, there may be a slight velocity gradient from west to east. In order to fully confirm the rotation motion, we need higher angular resolution observations.
Nevertheless, the CH$_3$OH emission extends over a velocity range of about 10 km s$^{-1}$.
Since the internal structure of the core is not resolved, the velocity structures are diluted in one beam. Thus, this relatively large velocity range may come from rotation or infall.

If the large velocity width would be due to only rotational motion and we assume Keplerian rotation, the stellar mass is roughly estimated to be about 11 $M_{\odot}$ and 27 $M_{\odot}$ for the distance of 1.56 kpc and 3.9 kpc, respectively; the rotational velocity, $v_{rot}$, is assumed to be 5 km s$^{-1}$ and the radius, $R$, is assumed to be 400 au and 950 au for the distance of 1.56 kpc and 3.9 kpc, respectively, and the mass is derived by $v_{rot}^2 R/G$, where $G$ is the gravitational constant.  

In the case of infall motion, the stellar mass is given by $v_{inf}^2 R/(2G)$, which is derived from $m$$v_{inf}^2$/2$-$$GmM$/$R$ = 0, assuming that the infall velocity ($v_{inf}$) at $R$=$\infty$ is 0.  In this case, the stellar mass is evaluated to be about 6 $M_{\odot}$ and 13 $M_{\odot}$ for the distance of 1.56 kpc and 3.9 kpc, respectively.
Since the emitting region that has such a large linewidth yet remains unresolved, these estimates can be recognized as upper limits. Furthermore, no luminous cm-source has been found toward this hot core in the NVSS data (Condon et al. 1998), which may suggest a lower mass.

In Figure 4a, we show the spectra of CH$_3$OH, HCOOCH$_3$ and CH$_3$CH$_2$CN toward the peak position of the CH$_3$OH $J_K$=10$_2$--9$_3$ $A^-$ line.
The CH$_3$OH line is found to be broader than the other molecular lines; it is extended toward higher velocity.
Although the S/N ratio may not be ideal for a comparison, we found that the HCOOCH$_3$ and CH$_3$CH$_2$CN emission are both detected in the same velocity range (53-62 km$^{-1}$). Since the spatial distribution of the integrated intensity is similar (Figure 1), the N-bearing and O-bearing COMs, except for CH$_3$OH, seems to have similar distributions in G34.43 MM3 HC.

In Figure 4b, we show the spectra of the deuterated molecules toward the same position as Figure 4a.  We can see that the peak velocity and the line shape of DNC is different from those of D$_2$CO and CH$_2$DCN.
Although the S/N ratio is poor, the line shapes of D$_2$CO and CH$_2$DCN look similar to that of CH$_3$CH$_2$CN, where the peak velocity of D$_2$CO and CH$_2$DCN is offset to the lower velocity than the peak velocity of the DNC line.
Hence, the D$_2$CO and CH$_2$DCN emission is likely to come from G34.43 MM3 HC.  On the other hand, the DNC line is narrower than the D$_2$CO and CH$_2$DCN lines, and the peak velocity of the DNC line is higher than those of the D$_2$CO and CH$_2$DCN lines. 
Since the DNC integrated intensity peak is offset form G34.43 MM3 HC, the DNC emission is likely to trace the envelope. In order to confirm this, higher angular resolution observations with better sensitivity are required.

\subsection{Column Densities}

We evaluate the column density of the observed molecules toward the peak of the CH$_3$OH $J_K$=10$_2$--9$_3$ $A^-$ emission from their integrated intensity, assuming the local thermodynamic equilibrium (LTE) condition.
The velocity range for the integration is $\pm$5 km s$^{-1}$ centered at the LSR velocity of 59.7 km s$^{-1}$.
For these column density estimates, we used the transition lines presented in Figure 1.
For the CH$_3$CH$_2$CN, HCOOCH$_3$, CH$_2$DCN, CH$_3$OCH$_3$ lines, the velocity range for the integration is set from -5 km s$^{-1}$ of the lowest velocity line to +5 km s$^{-1}$ of the highest velocity line to cover all the overlapping lines, {as mentioned in Section 3.1}. The velocity ranges for the integration are listed in Table 2.
The equations used for the column density derivation are presented in the Appendix. 

Since several CH$_3$OH lines are detected, we apply the rotation diagram method (Turner 1991; Goldsmith \& Langer 1999); we fit the data to the following equation;
\begin{equation}
{\rm log}\frac{3kW}{8\pi^3 \nu \mu^2 S} = {\rm log}\frac{N}{Q_{\rm rot}}-\frac{E_u}{k}\frac{{\rm log} e}{T_{\rm rot}},
\end{equation}
where $k$ is the Boltzmann constant, $W$ is the integrated intensity, $\nu$ is the rest frequency, $\mu$ is the electric dipole moment, $S$ is the line strength, $N$ is the column density, $Q$ is the partition function, $E_u$ is the upper state energy, and $T_{\rm rot}$ is the rotation temperature. Figure 5 shows the rotation diagram.

In Figure 5, considerable scatter is seen in the data.
This scattering could be partly due to optical depth effects; some transitions may be reached the optically thick limit, while others may still be optically thin (Goldsmith \& Langer 1999; Bisschop et al. 2007).
We estimate the optical depth of each line by using the RADEX code (van der Tak et al. 2007).
When we set the density of 10$^7$ cm$^{-3}$, the column density of 10$^{17}$ cm$^{-2}$ and the velocity width of 5 km s$^{-1}$, the optical depth of the CH$_3$OH $J_K$=5$_0$-4$_0$ $E^+$, $J_K$=5$_1$-4$_1$ $E^-$, and $J_K$=5$_0$-4$_0$ $A^+$ lines is calculated to be higher than unity. 
Hence, we exclude those lines (black circles in Figure 5) for the temperature estimate using the rotation diagram method.
In addition, we exclude the data of CH$_3$OH $J_K$=9$_{-1}$-8$_{0}$ $E$, which is a maser transition (Yanagida et al. 2014).
The data used for the fit are indicated by red circles in Figure 5.

As a result, the rotation temperature is derived to be 110$_{-20}^{+30}$ K.
We use this temperature for the derivation of the column density of all the observed molecules, including CH$_3$OH.
Although two or more different transition lines are detected for some of the other molecules, we use the brightest one for the derivation of the column density. 
The results are shown in Table 4.  The errors in Table 4 include the uncertainty of the excitation temperature and the 1 sigma rms noise level of the integrated intensity.

\section{DISCUSSION}

\subsection{Complex Organic Molecules}

\subsubsection{Comparison with the Other Sources}

To investigate the nature of G34.43 MM3 HC, we compare the molecular abundances in this source with those in the other low-mass and high-mass star-forming regions reported in the literature.
For comparison, we use the abundances relative to CH$_3$OH, 
because CH$_3$OH is thought to be the parent species in the synthesis of more complex organic molecules (e.g., Herbst \& van Dishoeck 2009).

Recently, Taquet et al. (2015) derived the averaged values of the abundances of several COMs relative to CH$_3$OH for hot corinos (bolometric luminosity ($L_{bol}$) $<$ 100 $L_{\odot}$) and high-mass hot cores ($L_{bol}$ $>$ 10$^4$ $L_{\odot}$). They found that the HCOOCH$_3$ and CH$_3$OCH$_3$ abundances relative to CH$_3$OH are comparable between low- and high-mass sources, while the CH$_3$CH$_2$CN abundance relative to CH$_3$OH is different.
In Table A10 of Taquet et al. (2015), all of CH$_2$CH$_3$CN/CH$_3$OH of high-mass sources, 
except for the upper limit data, is higher by more than one order magnitude than that of the low-mass sources. Thus, the difference in CH$_2$CH$_3$CN/CH$_3$OH between low-mass and high-mass sources is significant.
We compare our results of HCOOCH$_3$, CH$_3$OCH$_3$ and CH$_3$CH$_2$CN with their results (Figure 6).
In Figure 6, the abundances of all the three COMs relative to CH$_3$OH of G34.43 MM3 HC are similar to the average values of the high-mass hot cores.
The CH$_3$CH$_2$CN/CH$_3$OH ratio of G34.43 MM3 HC is higher by about one order of magnitude than that of the low-mass protostars.

Strong emission of the N-bearing COMs, such as CH$_3$CH$_2$CN, is often observed toward massive hot cores (e.g., Blake et al. 1996; Beuther et al. 2007; Qin et al. 2010), while it is weak toward low-mass sources (e.g., Jaber et al. 2014; Taquet et al. 2015). Furthermore, there is some evidence that N-bearing COMs are less abundant in intermediate-mass hot cores. For example, Fuente et al. (2014) suggest that the N-bearing COMs are not abundant in the intermediate-mass hot core NGC7129-FIR 2, and Palau et al. (2011) also suggest that they are not abundant in the intermediate-mass hot core IRAS 22198+6336 and AFGL 5124.
Thus, the abundance of the N-bearing COMs relative to CH$_3$OH in G34.43 MM3 HC seems to be similar to that of the high-mass hot cores rather than that of the hot corinos and intermediate-mass hot cores.

We compare our results with the COMs abundances relative to CH$_3$OH in four individual high-mass sources, which were also observed with interferometers (Figure 7). The spatial resolution of the observations of these four sources is 0.6-2.7 times different from that of our observations, whereas the spatial resolution of other single-dish observations is about 10 times different from those of our observations. Thus, we avoid the comparison of our data with the single-dish results in this plot.
In Figure 7, we also plot the data of the hot corinos in NGC 1333 IRAS 2A (Taquet et al. 2015).

In Figure 7, we can see that the COMs abundances relative to CH$_3$OH of G34.43 MM3 HC are comparable to those of G34.26+0.15 SE (Mookerjea et al. 2007) within the 1 sigma error.
G34.26+0.15 SE is located in the vicinity of three compact HII regions (Mookerjea et al. 2007), and the luminosity of G34.26+0.15 region ($L_{\rm FIR}$ = $\sim$5$\times$10$^5$ $L_{\odot}$: Churchwell et al. 1990) is much higher than that of G34.43+00.24 MM3 region (9$\times$10$^3$ $L_{\odot}$; Rathborne et al. 2005). Hence, G34.26+0.15 is a more active star-forming region than G34.43+00.24 MM3.
This comparison may indicate that the luminosity of nearby stars does not always affect the COMs abundances relative to CH$_3$OH in individual hot cores.

\subsubsection{Origins of the COMs}

COMs can be formed via three different era and/or processes: grain surface reactions in cold starless stage, gas-phase and grain-surface reactions in the warm up phase, and gas-phase reaction after the ice sublimation.
We have found that the HCOOCH$_3$ and CH$_3$OCH$_3$ abundances relative to CH$_3$OH toward G34.43 MM3 HC are similar to the averaged values of the other high-mass and low-mass sources.
The HCOOCH$_3$ and CH$_3$OCH$_3$ molecules are thought to be formed on grain surface with the reactions of CH$_3$O and HCO radicals (e.g. Garrod 2013).
Since such reactions occur efficiently during the warm-up phase with a temperature above $\sim$15 K, HCOOCH$_3$ and CH$_3$OCH$_3$ are thought to be formed mainly after the onset of star formation. 
Taquet et al. (2015) suggested that there is no clear trend between the luminosity of protostar and the abundances of HCOOCH$_3$ and CH$_3$OCH$_3$ relative to CH$_3$OH.
Our results confirm their suggestion.

On the other hand, the CH$_3$CH$_2$CN abundance relative to CH$_3$OH of G34.43 MM3 HC is higher than that of the hot corinos, and is comparable to that of the high-mass hot cores.
The CH$_3$CH$_2$CN molecule is thought to be formed on the grain surface by hydrogenations of C$_3$N or HC$_3$N (e.g. Caselli et al. 1993; Garrod et al. 2017).
Such hydrogenation may occur even in the cold ($\sim$10 K) phase.
However, if the CH$_3$CH$_2$CN molecule were formed mainly in the cold starless phase, it would be difficult to explain the observational results.
The abundance of the COMs on the grain surface generally increases with time, so that, if the CH$_3$CH$_2$CN molecule were formed in the cold starless phase, its abundance would depend on the timescale of the cold starless phase.
Since the timescale of the cold starless phase is determined by the infall and/or dissipation of turbulence and magnetic fields, it is not apparent that starless phase is systematically shorter in low-mass cores than in high-mass cores. In fact, variations of the starless phase timescale are suggested among low-mass cores (Hirota et al. 2002; Tafalla \& Santiago 2004; Sakai \& Yamamoto 2013), and high-mass cores might have some variations, as well (Kong et al. 2016). If so, the CH$_3$CH$_2$CN abundance of the low-mass sources should be as high as those of the high-mass sources in this picture.

Alternatively, Garrod et al. (2017) suggested from the chemical model calculations that the CH$_3$CH$_2$CN abundance on the grain surface increases by more than two order of magnitude after the onset of star formation.
Garrod et al. (2017) calculated the time dependence of chemical compositions in an accreting core with three different warm-up speeds (slow, medium, and fast) from 8 K to 400 K. According to Garrod et al. (2017), the CH$_3$CH$_2$CN/CH$_3$OH ratio is lower in the fast warm-up case ((0.3$\sim$1)$\times$10$^{-3}$) than in the medium ((4$\sim$5)$\times$10$^{-3}$) and slow warm-up cases ((1$\sim$3)$\times$10$^{-3}$) at the temperature of $\sim$130 K. Since the timescale of high-mass star formation is thought to be shorter than that of low-mass star formation (e.g. Viti \& Williams 1999), the observational results are apparently inconsistent with the model calculation results.
However, warm regions of a high-mass protostellar core is much larger than that of a low-mass protostellar core, and consequently, materials in a high-mass protostellar core could pass through warm regions for a longer time than those in low-mass protostellar cores (Aikawa et al. 2008; 2012). Thus, dust in high-mass protostellar cores could be warmed up slower than that in a low-mass protostellar cores.

The CH$_3$CH$_2$CN molecule can be also formed in the gas phase after the ice sublimation in the hot core phase ($T$ $>$ 100 K).
However, the formation timescale in the gas phase could be as long as 10$^5$ yr with a H$_2$ density of 10$^7$ cm$^{-3}$ and a cosmic-ray ionization rate of 10$^{-17}$ s$^{-1}$ (Rodgers \& Charnley 2001).
This is much longer than the outflow age of G34.43+00.24 MM3 ($<$1900 yr).  
If we assume that the embedded protostar significantly increased its radiation feedback, heating up the surrounding gas, at the moment of launching the outflow, we can say that the contribution of the gas-phase reaction seems to be negligible for the production of CH$_3$CH$_2$CN.
Moreover, the CH$_3$CH$_2$CN emission is also detected toward another young hot core in W3(H$_2$O) (Qin et al. 2015), which has outflows with a dynamical age of about 5$\times$10$^3$ yr (Qin et al. 2016), and the CH$_3$CH$_2$CN/CH$_3$OH ratio of W3(H$_2$O) is comparable to that of G34.43 MM3 HC.
It seems likely that the CH$_3$CH$_2$CN molecule could be formed on ice mantle during the warm-up phase.
The similar distribution between O-bearing and N-bearing COMs in G34.43 MM3 HC may be explained if both types of COMs have recently sublimated from ice mantles in the hot core.
However, this is just a single case study. Survey observations toward young hot cores/corinos are crucial in order to confirm the origin of CH$_3$CH$_2$CN.

\subsection{D$_2$CO}

\subsubsection{Comparison with the Other Sources}

Although interferometric observations of D$_2$CO toward hot cores/corinos are very limited, we compare the D$_2$CO abundance relative to CH$_3$OH (D$_2$CO/CH$_3$OH ratio) with that of some other sources (Figure 8).
Fuente et al. (2014) observed the D$_2$CO emission toward the intermediate-mass hot core NGC7129 FIR 2 with the PdBI; the dynamical age of the outflow associated with NGC7129 FIR 2 is $>$3$\times$10$^3$ yr (Fuente et al. 2001).
They found that the D$_2$CO/CH$_3$OH ratio of NGC7129 FIR 2 ($\sim$10$^{-4}$) is higher than that of Orion-KL ($\sim$4$\times$10$^{-5}$).
The D$_2$CO/CH$_3$OH ratio of G34.43 MM3 HC is evaluated to be 1.5$_{-1.0}^{+3.1}$$\times$10$^{-3}$, which is even higher than that of NGC7129 FIR 2.
Thus, the D$_2$CO abundance of G34.43 MM3 HC seems to be high, as compared with that of the other active hot cores.
In fact, the 4$_{0,4}$--3$_{0,3}$ emission of D$_2$CO, which is observed in this study, was not detected in the single-dish line surveys toward Orion-KL (Sutton et al. 1985) and Sgr BN (Nummelin et al. 1998), although the COMs emission is generally strong there.

Parise et al. (2006) observed D$_2$CO and CH$_3$OH lines toward several low-mass protostars with a single-dish telescope. 
According to their results, the D$_2$CO/CH$_3$OH ratio is 0.004-0.07 toward the low-mass protostellar cores, which is higher than that of G34.43 MM3 HC.
Since the observation by Parise et al. (2006) is done with the single-dish telescope, a part of the D$_2$CO emission may also come from a cold envelope; Parise et al. (2006) assumed the source size of 10$^{\prime\prime}$ in the derivations of the abundances, while the size of the hot corinos is at least 10 times smaller than the source size (Maret et al. 2004).　If so, the D$_2$CO/CH$_3$OH ratio of the hot corinos could be lower than the values reported for low-mass star forming cores. 
In fact, Persson et al. (2018) recently observed D$_2$CO toward the low-mass source IRAS 16293-2422 B with a resolution of 0$^{\prime\prime}$.5 with ALMA, and the D$_2$CO/CH$_3$OH ratio is found to be 8$_{-2.5}^{+2}$$\times$10$^{-4}$ (the CH$_3$OH data is taken from J{\o}rgensen et al. 2016).
Thus, G34.43 MM3 HC seems to have the similar D$_2$CO/CH$_3$OH ratio to that of hot corinos (Figure 8), although the abundance of COMs in G34.43 MM3 HC is similar to that of high-mass hot cores.

\subsubsection{Origin of D$_2$CO}

D$_2$CO is thought to be formed in the cold ($\sim$10 K) starless phase (Bacmann et al. 2003), by gas phase reactions and grain surface reactions (Hidaka et al. 2009). 
Once the core gets warm, deuterium fractionation becomes inefficient, while icy D$_2$CO sublimates.
The destruction timescale of D$_2$CO in the gas phase in a hot core is about 10$^5$ yr (Rodgers \& Millar 1996). Thus the D$_2$CO abundance at the onset of star formation could be retained after the protostar turns on.
The high abundance of D$_2$CO in G34.43 MM3 HC indicates that G34.43 MM3 HC is young.

At low temperatures ($\lesssim$ 20 K), molecular D/H ratio, including D$_2$CO/H$_2$CO (Taquet et al. 2012), increases with time.
Kong et al. (2016) calculated the time dependence of deuterium fractionation ratios with different collapsing speeds, and they found that the deuterium fractionation ratios at a given density depend on the timescale of starless phase.
Thus, the D$_2$CO abundance could depend on the timescale of its cold starless phase, where longer cold starless phase leads to higher D$_2$CO abundance.
If so, the high abundance of D$_2$CO in G34.43 MM3 HC would suggest a longer timescale of the cold ($<$ 20 K) starless phase than that of other hot cores.
However, the number of the D$_2$CO observations toward hot cores is still limited. To investigate whether there is diversity of the initial D$_2$CO abundance in hot cores, we need survey observations of D$_2$CO toward many sources in early evolutionary stages of star formation.

We may also say with the currently available data that G34.43 MM3 HC is a highly deuterated hot core/corino, although we need to derive the deuterium fractionation ratios, such as D$_2$CO/H$_2$CO, for a definitive conclusion. 
We detected the CH$_2$DCN emission toward G34.43 MM3 HC.
The reports of the detection of CH$_2$DCN are very limited (Sgr B2(N2): Belloche et al. 2016; Orion-KL and G34: Gerin et al. 1992).  It is possible that the CH$_2$DCN abundance is also high toward G34.43 MM3 HC, while the derivation of the deuterium fractionation ratios, CH$_2$DCN/CH$_3$CN, is essential.

Finally, we should note that the DNC emission is weak toward the hot core; i.e. the peak position of DNC emission is offset from G34.43 MM3 HC. According to the chemical model calculations (Sakai et al. 2015), the DNC abundance in the ice mantle is higher than that in the gas phase before the onset of star formation, and thus the DNC emission should peak toward G34.43 MM3 HC, once DNC sublimates in the hot core (the sublimation temperature is about 80 K). Sakai et al. (2015) pointed out the possibility that the size of G34.43 MM3 HC is very small and the beam filling factor of the hot core is low even with the 0$^{\prime\prime}$.8 beam. However, the D$_2$CO emission is clearly detected to have a peak toward G34.43 MM3 HC.
The size of the hot core is thus large enough to be detected with emission lines of sublimated molecules.
It implies that DNC is destroyed more efficiently than D$_2$CO in the hot core. Hirota et al. (1998) observationally suggest that the HNC/HCN ratio decreases with increasing temperature, due to the reaction of HNC + H. This reaction is also consider to be responsible for the decline of the HNC/HCN ratio at high temperatures in chemical models (Schilke et al. 1992; Graninger et al. 2014).
It should be noted, however, that this reaction destroy HNC, as well, while the HN$^{13}$C emission peaks toward the hot core in our observation (Sakai et al. 2015). Further theoretical studies, as well as observations with higher angular resolution are desirable to understand the evolution of HNC and DNC around the hot core.

\subsection{Nature of the Hot Core/Corino in G34.43+00.24 MM3}

We have shown that the COM abundances relative to CH$_3$OH of G34.43 MM3 HC are similar to those of high-mass hot cores, rather than those of hot corinos. 
On the other hand, the D$_2$CO abundance relative to CH$_3$OH of G34.43 MM3 HC is estimated to be similar to that of hot corinos, rather than that of high-mass hot cores.
Thus, G34.43 MM3 HC seems to have an unique and interesting chemical composition.

The similarity of the COM abundances relative to CH$_3$OH between G34.43 MM3 HC and the other high-mass hot cores indicates that G34.43 MM3 HC is not an assembly of several unresolved hot corinos, although it is still unclear whether G34.43 MM3 HC consists of a single core or not.
The similar COMs abundances relative to CH$_3$OH may also suggest that the physical conditions during the warm-up phase of G34.43 MM3 HC are similar to those of the other high-mass hot cores, because the COMs are thought to be mainly formed in the warm-up phase, as pointed out in Section 4.1.2. 
Since the total mass of the clump is about 300 $M_\odot$ in the case of the distance of 3.7 kpc (Rathborne et al. 2006), it could be possible that the core will increase the mass and will form a high-mass star in the future. 
Indeed, several studies in IRDCs suggest that cores that will form high-mass stars begin their evolution in a low-mass regime (eg., Zhang et al. 2009, 2015; Ohashi et al. 2016; Sanhueza et al. 2017). The low-mass cores eventually accrete sufficient mass to form massive stars, resembling competitive accretion scenarios (Bonnell \& Bate 2006; Wang et al. 2010).

On the other hand, we have suggested from the observed D$_2$CO/CH$_3$OH ratio that the duration time of starless phase of G34.43 MM3 HC is longer than that of hot cores.
Although it is thought that the G34.43+00.24 MM3 clump has experienced low-mass star formation in the past (Sakai et al. 2013; Foster et al. 2014), this low-mass star formation event would not have been so intense to significantly affect G34.43 MM3 HC.
Consequently, the cold starless phase of G34.43 MM3 HC could be long.

In order to investigate the nature of G34.43 MM3 HC in more detail, we need to resolve the circumstellar structures around the protostar with higher angular resolution observations.
This source could be a good target in order to investigate not only the physical processes but also the chemistry in the early stage of star formation in a cluster-forming region.

\section{Summary}

We observed COMs and deuterated species toward G34.43+00.24 MM3 HC with ALMA.  The main results are summarized below.

\begin{itemize} 

\item We detected six normal-COMs (CH$_3$OH, CH$_3$CHO, CH$_3$CH$_2$CN, CH$_3$OCH$_3$, HCOOCH$_3$, and NH$_2$CHO), one deuterated-COM (CH$_2$DCN), and two deuterated fundamental molecules (D$_2$CO and DNC), above the 3 sigma rms noise level toward G34.43 MM3 HC. The distribution is similar between the N-bearing and O-bearing molecules.

\item We compared the abundances of COMs relative to CH$_3$OH with those of other regions. The COMs abundances relative to CH$_3$OH in G34.43 MM3 HC are similar to those in high-mass hot cores, rather than those in hot corinos and intermediate-mass hot cores. We suggest that the COMs are mainly formed on ice mantles during the warm-up phase, and that the physical conditions of the warm-up phase of G34.43 MM3 HC are similar to that of other high-mass hot cores.

\item The distribution of DNC is different from that of D$_2$CO and CH$_2$DCN. The D$_2$CO and CH$_2$DCN emission likely comes from G34.43 MM3 HC, while the DNC emission could trace the outer colder envelope around G34.43 MM3 HC.

\item The D$_2$CO abundance relative to CH$_3$OH is relatively high toward G34.43 MM3 HC. 
This indicates that the G34.43 MM3 HC is young, and may also suggest that the cold starless phase of G34.43 MM3 HC has lasted longer than that in other hot cores.

\end{itemize} 

Detailed molecular line observations of hot cores in the IRDCs are still limited.
The observations of the early stages of the hot core phase are important in order to investigate the formation mechanism of not only stars, but also molecules.
In order to investigate the origin of the COMs and deuterated molecules in hot cores, we need to achieve a spatial resolution of $\sim$several 100 au toward protostars (e.g. Oya et al. 2016).
By using ALMA, we will be able to conduct statistical studies with such a high-angular resolution.

\acknowledgments

We would like to thank the anonymous referee for useful comments that improved our manuscript.
This paper makes use of the following ALMA data: ADS/JAO.ALMA\#2011.0.00656.S. 
ALMA is a partnership of ESO (representing its member states), NSF (USA) and NINS (Japan), together with NRC (Canada) and NSC and ASIAA (Taiwan), in cooperation with the Republic of Chile. 
The Joint ALMA Observatory is operated by ESO, AUI/NRAO, and NAOJ. We are grateful to the ALMA staff.
This study is supported by KAKENHI (25400225, 16K05292 and 25108005).





\appendix
 
\section{Derivation of Column density}\label{sec:dcol}

We derive the column density assuming LTE and optically thin emission. The column density is derived by using the following equation;
\begin{equation}
N_{total}=\frac{3c^2}{16\pi^3 \nu^3 \mu_0^2 S \Omega_{\rm B}}Q\exp\left(\frac{E_u}{kT_{\rm ex}}\right) \left[1-\frac{J(T_{\rm BB})}{J(T_{\rm ex})}\right] \int S_\nu dv, \label{eq:col}
\end{equation}
where $c$ is the speed of light, $\nu$ is the rest frequency, $Q$ is the partition function, $\mu$ is the electric dipole moment, $S$ is the line strength, $\Omega_{\rm B}$ is the solid angle of the beam, $E_u$ is the upper state energy, $T_{\rm ex}$ is the excitation temperature, $S_\nu$ is the flux, $k$ is Boltzmann constant, and $J$ stands for the temperature given as
\begin{equation}
J(T) = \frac{\frac{h\nu}{k}}{\exp\left(\frac{h\nu}{kT}\right)-1}.
\end{equation}
In Equation A1, we use the following relation between the main beam temeprature ($T_{MB}$) and $S_\nu$,
\begin{equation}
T_{MB} = \frac{c^2}{2k\nu^2 \Omega_B}S_{\nu}.
\end{equation} 

The partition function $Q$ is evaluated from the data in JPL or CDMS. We fit the JPL or CDMS data to the function of $Q$($T$) = $a$$\times$$T^b$, where $a$ and $b$ are fitting parameters. The fitting results are listed in Table 5.




\clearpage




\begin{figure}
\epsscale{1.0}
\plotone{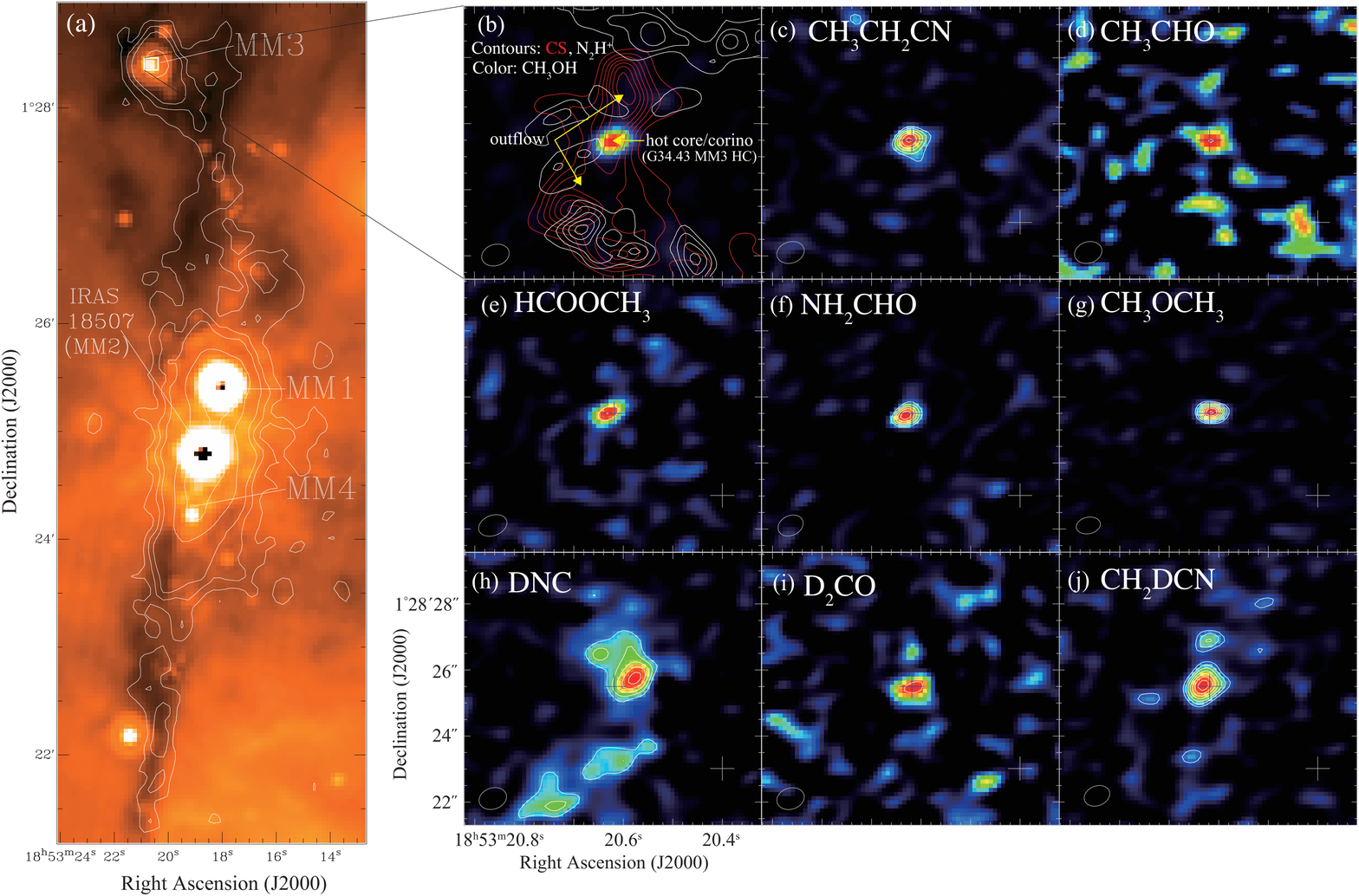}
\caption{(a) $Spitzer$ 24 $\mu$m color image overlaid with the 1.2 mm continuum (Rathborne et al. 2006). (b) Integrated intensity images of CH$_3$OH $J_K$=10$_2$--9$_3$ $A^-$ (color), N$_2$H$^+$ $J$=3--2 (white contours) and $^{13}$CS $J$=5--4 (red contours), (c) CH$_3$CH$_2$CN 27(0,27)--26(1,26), 24(2,23)--23(1,22), and 26(1,25)--(25(1,24), (d) CH$_3$CHO 12(5,8)--11(5,7) and 12(5,7)--11(5,6), (e) HCOOCH$_3$ 21(3,18)--20(3,17), 21(3,19)--20(3,18) $A$, 20(9,12)--19(3,11) $A$ and 29(9,11)--19(9,10) $A$, (f) NH$_2$CHO 12(1,11)--11(1,10), (g) CH$_3$OCH$_3$ 12(2,11)--11(1,10) EE and 12(2,11)--11(1,10) AA, (h) DNC $J$=3--2, (i) D$_2$CO 4(0,4)--3(0,3), and (j) CH$_2$DCN 15(2,14)--14(2,13), 15(3,13)--14(3,12), and 15(3,12)--14(3,11).
Grey and blue cross marks represent the position of the phase center and the peak of CH$_3$OH $J$=10$_2$--9$_3$ $A^-$ emission, respectively.
Contour levels start from the 3$\sigma$ noise level and increase in steps of 1$\sigma$ [(c) $1\sigma=20$ mJy beam$^{-1}$ km s$^{-1}$, (d) $1\sigma=20$ mJy beam$^{-1}$ km s$^{-1}$, (e) $1\sigma=20$ mJy beam$^{-1}$ km s$^{-1}$, (f) $1\sigma=20$ mJy beam$^{-1}$ km s$^{-1}$, (g) $1\sigma=30$ mJy beam$^{-1}$ km s$^{-1}$, (h) $1\sigma=10$ mJy beam$^{-1}$ km s$^{-1}$, (i) $1\sigma=10$ mJy beam$^{-1}$ km s$^{-1}$, and (j) $1\sigma=10$ mJy beam$^{-1}$ km s$^{-1}$]. \label{fig1}}
\end{figure}

\begin{figure}
\epsscale{1.0}
\plotone{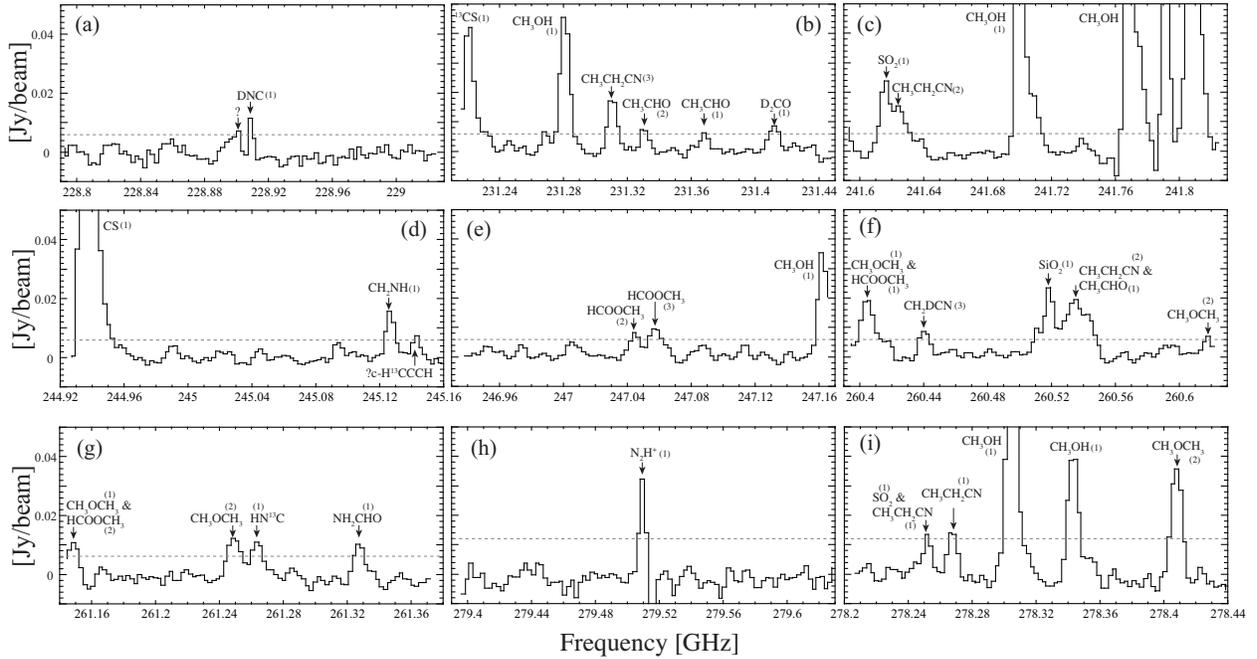}
\caption{Spectra toward the hot core/corino in G34.43+00.24 MM3. The number in parentheses represents the number of the overlapping lines of the molecule. The dotted lines indicate the 3 sigma noise level per binned channel: $1\sigma=0.002$ Jy beam$^{-1}$ for (a)-(g) and $1\sigma=0.004$ Jy beam$^{-1}$ for (h) and (i). Channel width of the data used in this figure is 2.44 MHz.\label{fig2}}
\end{figure}


\begin{figure}
\epsscale{1.0}
\plotone{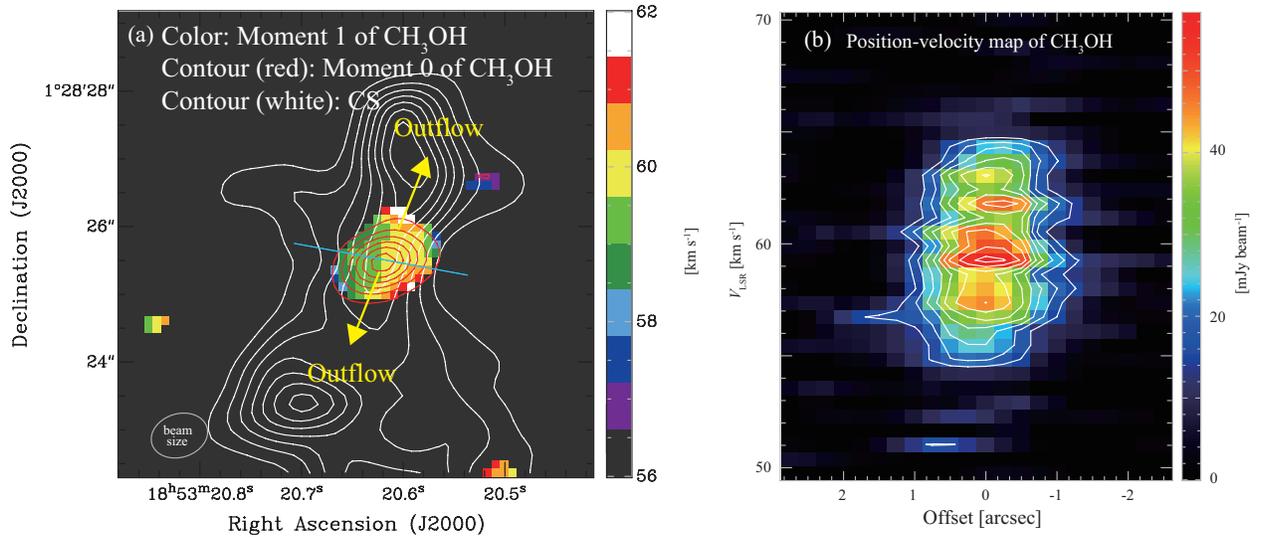}
\caption{(a) Moment 1 map of CH$_3$OH $J_K$=10$_2$--9$_3$ $A^-$ (color) toward G34.43+00.24 MM3 HC. The integrated intensity maps of CH$_3$OH $J_K$=10$_2$--9$_3$ $A^-$ (red contours) and CS $J$=5--4 (white contours) are superposed. The velocity range for the moment 1 map is 50-70 km s$^{-1}$. (b) Position-velocity diagram of CH$_3$OH $J_K$=10$_2$--9$_3$ $A^-$ through the cut indicated in (a).\label{fig3}}
\end{figure}

\begin{figure}
\epsscale{0.6}
\plotone{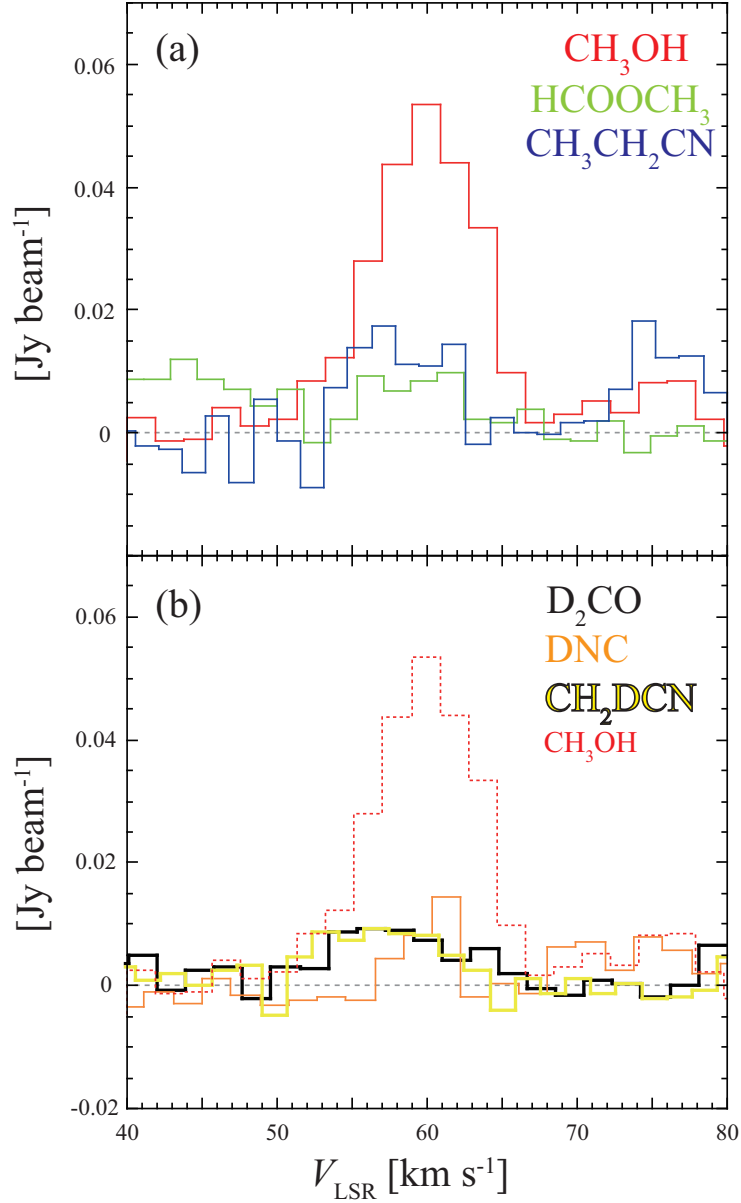}
\caption{(a) Spectra of CH$_3$OH $J_K$=10$_2$--9$_3$ $A^-$, HCOOCH$_3$ 21(3,18)--20(3,17) $E$ and CH$_3$CH$_2$CN 31(6,25)--30(6,24) toward G34.43+00.24 MM3 HC. (b) Spectra of D$_2$CO 4(0,4)--3(0,3), DNC $J$=3--2, CH$_2$DCN 15(2,14)-14(2,13), and CH$_3$OH $J_K$=10$_2$--9$_3$ $A^-$ toward the same position as (a). The 1$\sigma$ rms noise level per binned channel is 0.002 Jy/beam for the data used in (a) and (b), except for CH$_3$CH$_2$CN (0.004 Jy/beam). The channel width is 1.90 km s$^{-1}$, 1.78 km s$^{-1}$, 1.58 km s$^{-1}$, 1.90 km s$^{-1}$, 1.92 km s$^{-1}$, and 1.69 km s$^{-1}$ for CH$_3$OH, HCOOCH$_3$, CH$_3$CH$_2$CN, D$_2$CO, DNC, and CH$_2$DCN, respectively.\label{fig4}}
\end{figure}

\begin{figure}
\epsscale{1.0}
\plotone{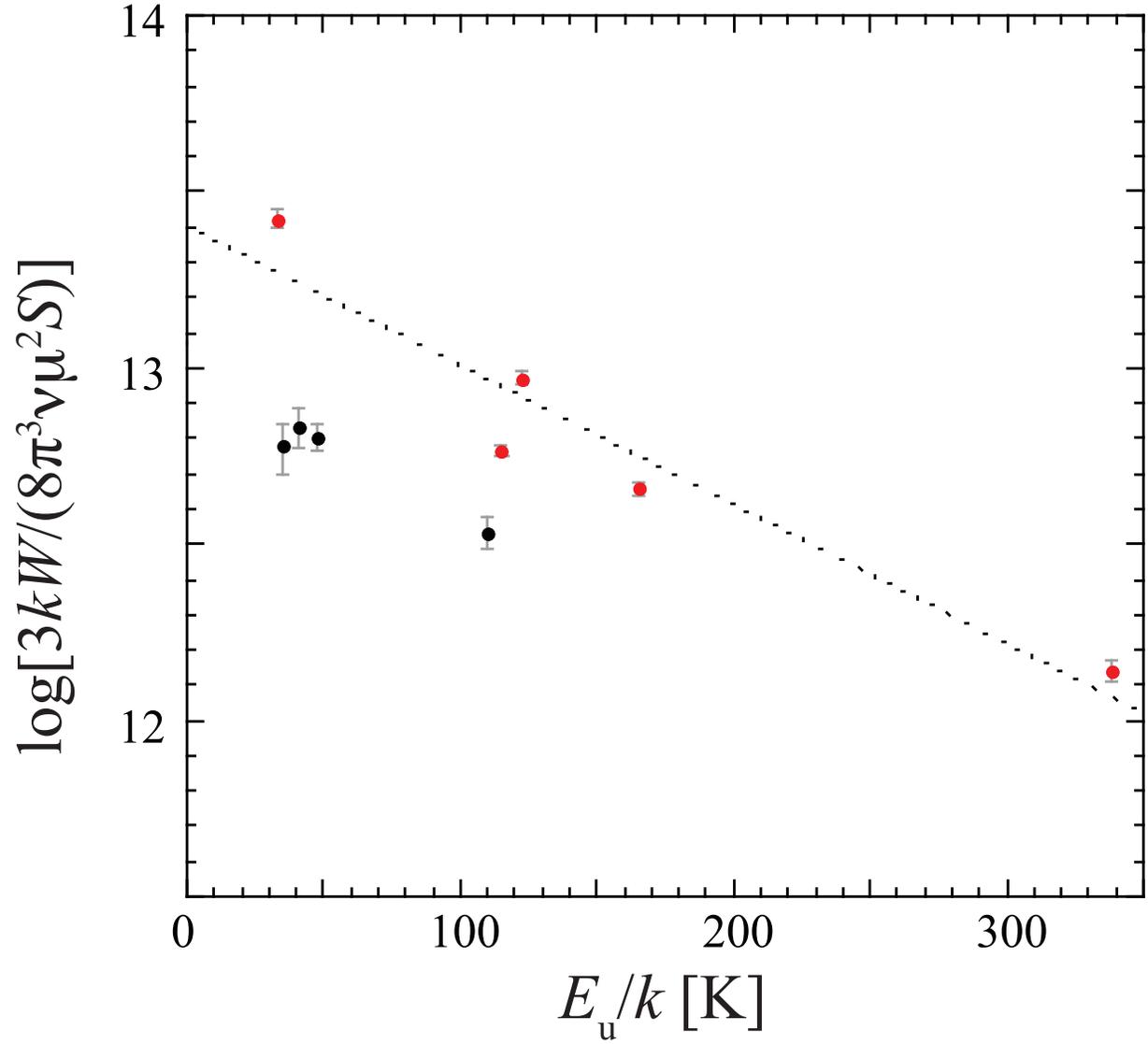}
\caption{Rotation diagram of the CH$_3$OH lines. Red circles indicate the data used for the fit, and black circles correspond to the neglected lines due to high optical depths (see also the text). The dotted line represents the best fit.\label{fig5}}
\end{figure}

\begin{figure}
\epsscale{1.0}
\plotone{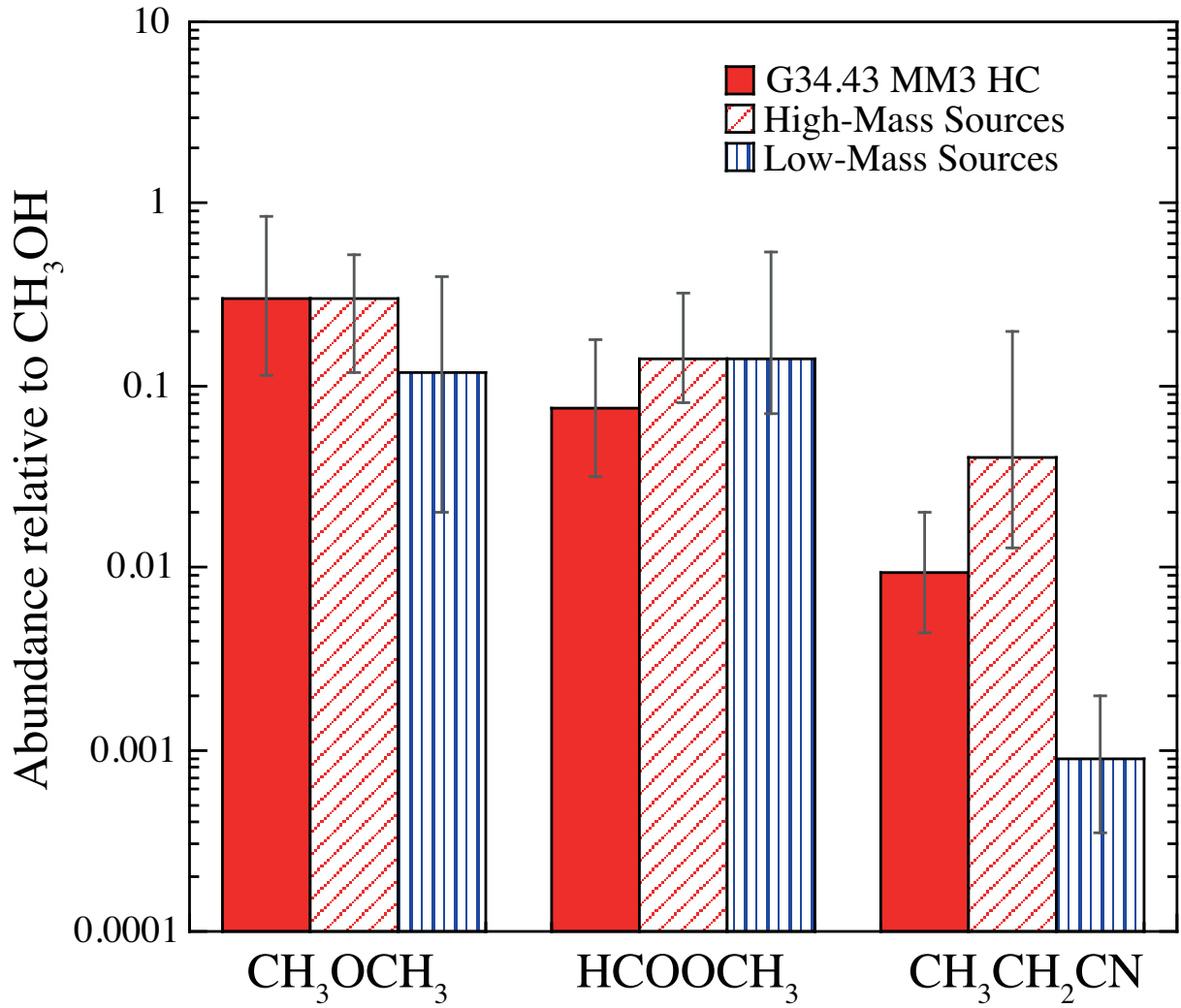}
\caption{COMs abundances relative to CH$_3$OH for G34.43+00.24 MM3 HC and averaged values of low-mass star forming regions and high-mass hot cores (Taquet et al. 2015)\label{fig6}}
\end{figure}

\begin{figure}
\epsscale{1.0}
\plotone{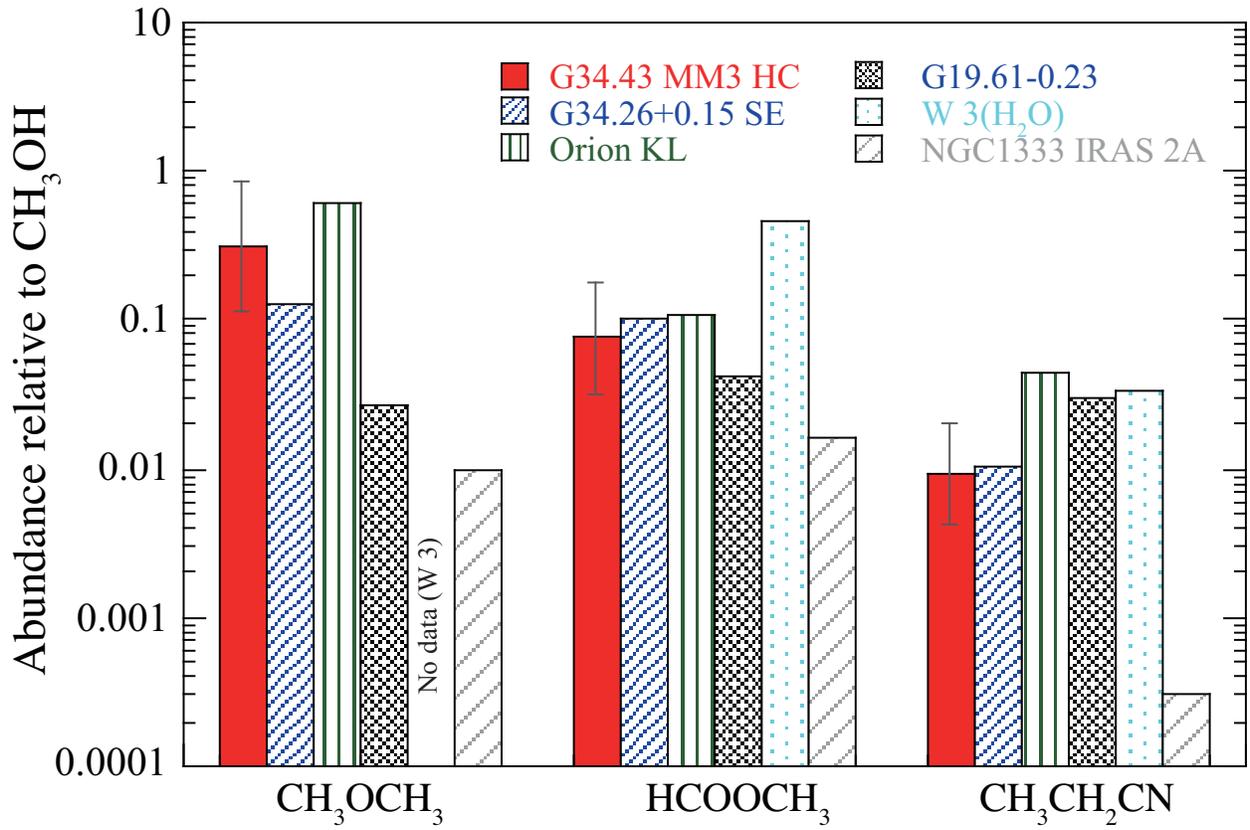}
\caption{COMs abundances relative to CH$_3$OH for G34.43+00.24 MM3 HC and other sources: G34.26+0.15 SE (Mookerjea et al. 2007), Orion KL (Feng et al. 2015), G19.61-0.23 (Qin et al. 2010), W 3(H$_2$O) (Qin et al. 2015) and NGC 1333 IRAS 2A (Taquet et al. 2015).\label{fig7} }
\end{figure}

\begin{figure}
\epsscale{1.0}
\plotone{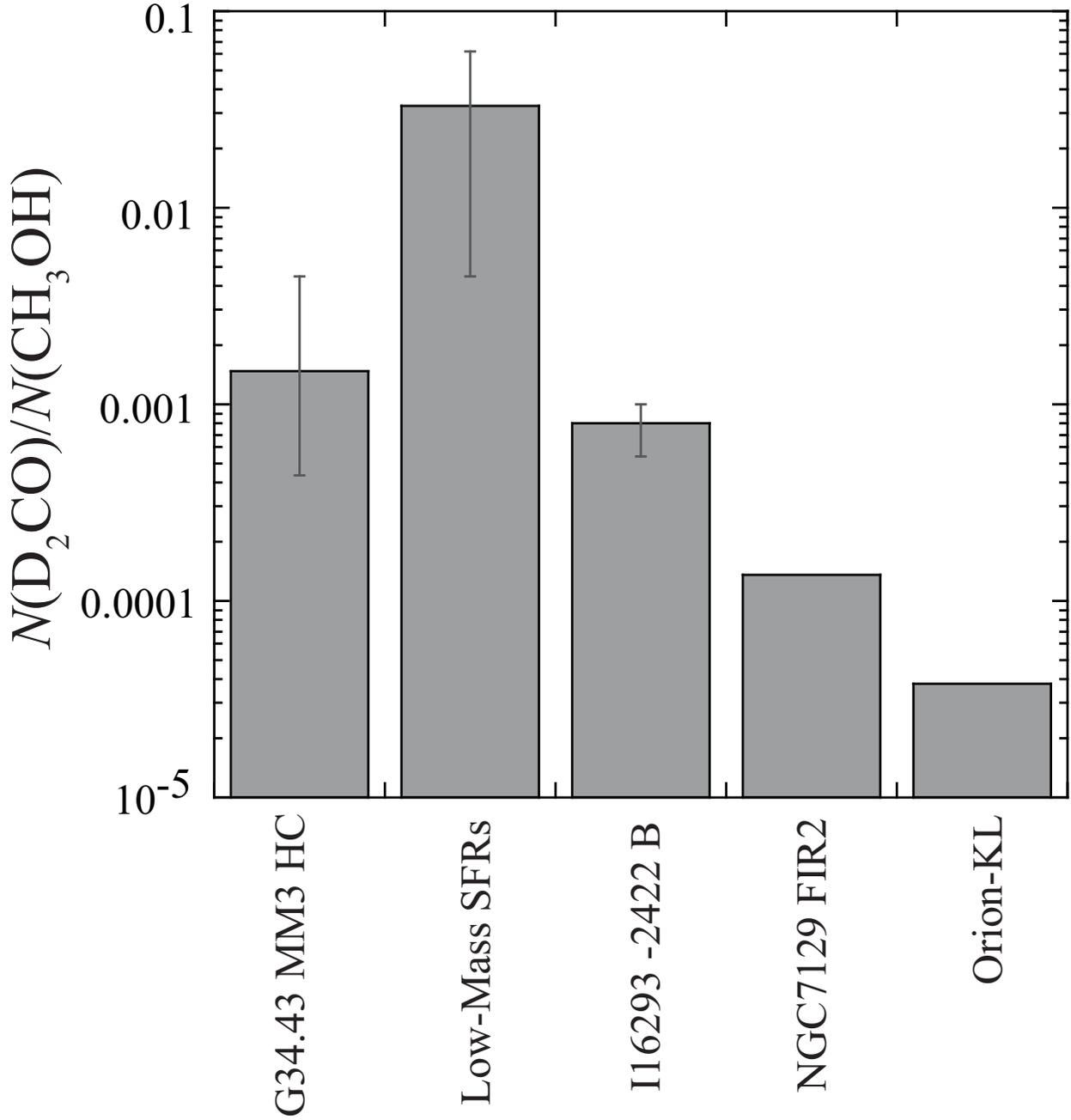}
\caption{D$_2$CO abundances relative to CH$_3$OH of G34.43+00.24 MM3 HC, the averaged value of low-mass star forming regions (Parise et al. 2006), IRAS 16293-2422 B (J{\o}rgensen et al. 2016; Persson et al. 2018), NGC7129 FIR 2 (Fuente et al. 2014), and Orion KL (Fuente et al. 2014 and references therein). \label{fig8}}
\end{figure}







\clearpage

\begin{deluxetable}{rrrrr}
\tabletypesize{\scriptsize}
\tablecaption{Frequency ranges for the 9 spectral windows\label{tbl-1}}
\tablewidth{0pt}
\tablehead{
\colhead{Window} & \colhead{Frequency\tablenotemark{a}} & \colhead{Beam Size} &  \colhead{P.A.} & \colhead{rms\tablenotemark{b}} \\
 &  \footnotesize[GHz]  &  & \footnotesize[degree] & \footnotesize[mJy beam$^{-1}$]\\
}
\startdata
1 & 228.793--289.025 & 0.85$^{\prime\prime}$$\times$0.64$^{\prime\prime}$ & -72 & 1.7 \\
2 & 231.216--231.448 & 0.85$^{\prime\prime}$$\times$0.64$^{\prime\prime}$ & -73 & 2.5 \\
3 & 241.593--241.825 & 0.79$^{\prime\prime}$$\times$0.60$^{\prime\prime}$ & 101 & 2.1 \\
4 & 244.932--245.166 & 0.78$^{\prime\prime}$$\times$0.59$^{\prime\prime}$ & 101 & 2.2 \\
5 & 246.937--247.166 & 0.89$^{\prime\prime}$$\times$0.61$^{\prime\prime}$ & -69 & 2.3 \\
6 & 260.394--260.623 & 0.80$^{\prime\prime}$$\times$0.59$^{\prime\prime}$ & 112 & 2.0 \\
7 & 261.144--261.372 & 0.79$^{\prime\prime}$$\times$0.60$^{\prime\prime}$ & -65 & 2.4 \\
8 & 278.206--278.440 & 0.74$^{\prime\prime}$$\times$0.50$^{\prime\prime}$ & 101 & 4.2 \\
9 & 279.395--279.629 & 0.73$^{\prime\prime}$$\times$0.49$^{\prime\prime}$ & 101 & 4.3 \\
\enddata
\tablenotetext{a}{$V_{LSR}$ = 59.7 km s$^{-1}$.}
\tablenotetext{b}{The rms noise level is derived from the channels free of emission in the cube with the channel width of 1.464 MHz.}
\end{deluxetable}

\clearpage

\begin{deluxetable}{ccrrrrr}
\tabletypesize{\scriptsize}
\tablecaption{Detected lines\label{tbl-1}}
\tablewidth{0pt}
\tablehead{
\colhead{Species} & \colhead{Transition} & \colhead{Frequency\tablenotemark{a}} & \colhead{$E_u/k$\tablenotemark{a}} & \colhead{$\mu^2 S$\tablenotemark{a}} & \colhead{Integrated Intensity} & \colhead{Velocity range\tablenotemark{b}}   \\
 & & \footnotesize[GHz]  & \footnotesize[K] &  \footnotesize[D$^2$] & \footnotesize[mJy beam$^{-1}$ km s$^{-1}$] & \footnotesize[km s$^{-1}$]\\
}
\startdata
 & & Window 1 & & & & \\
U-line & --- &  &  & & & \\
DNC & 3--2 & 228.91049 & 22.0 & 27.9 & 50$\pm$10 & 10 \\
 & &Window 2 & & & &\\
$^{13}$CS  & 4(1,4)--3(1,3) & 231.22069 & 33.3 & 38.3 & 370$\pm$20 & 10 \\
CH$_3$OH & 10$_2$--9$_3$ $A^-$ & 231.28110 & 165.3 & 2.8 & 400$\pm$20 & 10 \\
CH$_3$CH$_2$CN & 27(0,27)--26(1,26) & 231.31230 & 157.7 & 35.1 & 160$\pm$20 (blended) & 12 \\
CH$_3$CH$_2$CN & 24(2,23)--23(1,22) & 231.31324 & 132.4 & 18.4 &  blended &\\
CH$_3$CH$_2$CN & 26(1,25)--25(1,24) & 231.31042 & 153.4 & 383.1 & blended & \\
CH$_3$CHO & 12(5,8)--11(5,7) & 231.32970 & 128.6 & 125.5 & 60$\pm$20 (blended) & 10\\
CH$_3$CHO & 12(5,7)--11(5,6) & 231.32970 & 128.6 & 125.5 & blended &\\
CH$_3$CHO & 12(5,7)--11(5,6) & 231.36329 & 128.5 & 125.5 & $<$60 & 10\\
D$_2$CO & 4(0,4)--3(0,3) & 231.41023 & 27.9 & 43.4 & 70$\pm$20 & 10 \\
 & &Window 3 & & & & \\
SO$_2$ & 5(2,4)--4(1,3) & 241.61580 & 23.6 & 5.7 & 190$\pm$20 (blended) & 10\\
CH$_3$CH$_2$CN & 27(3,25)--26(3,24) & 241.62587 & 172.6 & 395.0 & blended &\\
CH$_3$OH & 5$_0$--4$_0$ $E^+$ & 241.70017 & 47.9 & 4.0 & 840$\pm$70 & 10 \\
CH$_3$OH & 5$_1$--4$_1$ $E^-$ & 241.76722 & 40.4 & 3.9 & 860$\pm$110 & 10\\
CH$_3$OH & 5$_0$--4$_0$ $A^+$ & 241.79143 & 34.8 & 4.0 & 790$\pm$130 & 10\\
CH$_3$OH & 5$_4$--4$_4$ $A^{\pm}$ & 241.80651 & 115.2 & 1.5 & 550$\pm$20 & 10\\
CH$_3$OH & 5$_4$--4$_4$ $E^-$ & 241.81326 & 122.7 & 1.4 & 440$\pm$20 & 10\\
 & &Window 4 & & & \\
CS & 5--4 & 244.93556 & 35.3 & 19.1 &  1820$\pm$110 & 10\\
CH$_2$NH & 4(1,4)--3(1,3) & 245.12587 & 37.3 & 20.2 & 120$\pm$10 & 10\\
U-line & --- &  &  & & &\\
& & Window 5  & & & & \\
HCOOCH$_3$ & 21(9,11)--20(9,10) E & 247.04065 & 177.8 & 42.5 & 80$\pm$20 (blended) & 14\\
HCOOCH$_3$ & 21(3,18)--20(3,17) E & 247.04408 & 139.9 & 53.9 & blended &\\
HCOOCH$_3$ & 21(3,19)--20(3,18) A & 247.05345 & 139.9 & 53.9 & 110$\pm$20 (blended) & 15\\
HCOOCH$_3$ & 20(9,12)--19(3,11) A & 247.05726 & 177.8 & 42.5 & blended & \\
HCOOCH$_3$ & 29(9,11)--19(9,10) A & 247.05774 & 177.8 & 42.5 & blended &\\
CH$_3$OH & 16$_2$--15$_3$ $E$ & 247.16199 & 338.1 & 4.8 & 270$\pm$20 & 10\\
& & Window 6 & & & & \\
CH$_3$OCH$_3$ & 16(5,11)-16(4,12) EE & 260.40339 & 159.0 & 177.1 & 170$\pm$20 (blended) & 10\\
HCOOCH$_3$ & 21(8,14)--20(8,13) E & 260.40403 & 179.2 & 47.6 & blended & \\
CH$_2$DCN & 15(2,14)--14(2,13) E & 260.43855 & 121.6 & 226.4 & 90$\pm$10 (blended) & 13\\
CH$_2$DCN & 15(3,13)--14(3,12) E & 260.44139 & 148.5 & 221.3 & blended & \\
CH$_2$DCN & 15(3,12)--14(3,11) E & 260.44197 & 148.5 & 221.3 & blended & \\
SiO & 6--5 & 260.51802 & 43.8 & 57.6 & 180$\pm$60 & 10 \\
CH$_3$CH$_2$CN & 29(5,25)--28(5,24) & 260.53569 & 215.1 & 417.0 &260$\pm$30 (blended) & 16\\
CH$_3$CHO & 11(4,8)--10(3,7) & 260.54114 & 46.2 & 7.0 & (blended) & \\
CH$_3$OCH$_3$  & 25(5,21)--25(4,22) AA & 260.61728 & 331.7 & 113.9 & 50$\pm$10 (blended) & 10\\
CH$_3$OCH$_3$  & 25(5,21)--25(4,22) EE & 260.61687 & 331.7 & 303.6 & blended & \\
 & & Window 7 & & & & \\
CH$_3$OCH$_3$ & 17(5,13)--17(4,14) EE & 261.14790 & 174.5 & 191.0 & 90$\pm$20 (blended) &12\\
HCOOCH$_3$ & 21(5,17)--20(5,16)  & 261.14890 & 154.1 & 52.6 &  blended & \\
CH$_3$OCH$_3$  & 15(5,10)--15(4,11) EE  & 261.24824 & 144.4 & 161.8 & 90$\pm$20 (blended) & 12\\
CH$_3$OCH$_3$  & 15(5,10)--15(4,11) AA & 261.25065 & 144.4 & 102.4 & blended & \\
HN$^{13}$C & 3--2 & 261.26348 & 25.1 & 21.8 & 100$\pm$20 & 10 \\
NH$_2$CHO & 12(1,11)--11(1,10) & 261.32745 & 84.9 & 155.6 & 130$\pm$20 & 10 \\
 & & Window 8  & & & & \\
N$_2$H$^+$ & 3--2  & 279.51173 & 26.8 & 335.2 & 90$\pm$70 & 10 \\
 & & Window 9 & & & & \\
SO$_2$ & 19(7,13)--20(6,14)  & 278.25097 & 294.8 & 6.1 & 90$\pm$20 (blended) & 10 \\
CH$_3$CH$_2$CN & 31(6,26)--30(6,25) & 278.25123 & 253.4 & 442.3 & blended & 10 \\
CH$_3$CH$_2$CN & 31(6,25)--30(6,24) & 278.26670 & 253.4 & 442.3 & 110$\pm$20 & 10 \\
CH$_3$OH & 9$_{-1}$--8$_0$ $E$ & 278.30451 & 110.0 & 5.8 & 340$\pm$20 & 10 \\
CH$_3$OH & 2$_{-1}$--3$_{-1}$ $E$ & 278.34226 & 32.9 & 3.3 & 780$\pm$80 & 10 \\
CH$_3$OCH$_3$  & 12(2,11)--11(1,10) EE & 278.40704 & 76.3 & 103.9 & 310$\pm$20 (blended) & 12 \\
CH$_3$OCH$_3$  & 12(2,11)--11(1,10) AA & 278.40874 & 76.3 & 64.9 & blended & \\
\enddata
\tablenotetext{a}{From CDMS(http://www.astro.uni-koeln.de/cdms) or JPL(http://spec.jpl.nasa.gov/)}
\tablenotetext{b}{Velocity range for the integrated intensity. For non-blended lines, the velocity range for the integration is $\pm$5 km s$^{-1}$ centered at the LSR velocity of 59.7 km s$^{-1}$. For blended lines, the velocity range for the integration is set from -5 km s$^{-1}$ of the lowest velocity line to +5 km s$^{-1}$ of the highest velocity line.  The 1$\sigma$ noise level is derived from the data of the emission free regions in the integrated intensity maps. }
\end{deluxetable}


\clearpage

\begin{deluxetable}{crrr}
\tabletypesize{\scriptsize}
\tablecaption{Results of the 2D Gaussian Fitting\tablenotemark{a}\label{tbl-1}}
\tablewidth{0pt}
\tablehead{
\colhead{Species} & \colhead{Source Size} & \colhead{Deconvolved Source Size}  & \colhead{Beam Size} \\}
\startdata
CH$_3$CH$_2$CN  & 0.95$^{\prime\prime}$($\pm$0.11)$\times$0.80$^{\prime\prime}$($\pm$0.08)  & 0.51$^{\prime\prime}$($\pm$0.25)$\times$0.41$^{\prime\prime}$($\pm$0.21) & 0.85$^{\prime\prime}$$\times$0.64$^{\prime\prime}$ \\
HCOOCH$_3$  & 1.01$^{\prime\prime}$($\pm$0.14)$\times$0.63$^{\prime\prime}$($\pm$0.06)& 0.48$^{\prime\prime}$($\pm$0.27)$\times$0.10$^{\prime\prime}$($\pm$0.20) & 0.89$^{\prime\prime}$$\times$0.61$^{\prime\prime}$ \\
NH$_2$CHO  & 0.83$^{\prime\prime}$($\pm$0.10)$\times$0.59$^{\prime\prime}$($\pm$0.05)& point source & 0.79$^{\prime\prime}$$\times$0.60$^{\prime\prime}$ \\
CH$_3$OCH$_3$  & 0.78$^{\prime\prime}$($\pm$0.06)$\times$0.51$^{\prime\prime}$($\pm$0.03) & point source & 0.73$^{\prime\prime}$$\times$0.49$^{\prime\prime}$\\
CH$_2$DCN & 1.09$^{\prime\prime}$($\pm$0.15)$\times$0.88$^{\prime\prime}$($\pm$0.10) & 0.78$^{\prime\prime}$($\pm$0.26)$\times$0.60$^{\prime\prime}$($\pm$0.35) & 0.80$^{\prime\prime}$$\times$0.59$^{\prime\prime}$\\
CH$_3$OH & 0.98$^{\prime\prime}$($\pm$0.05)$\times$0.67$^{\prime\prime}$($\pm$0.03)  & 0.49$^{\prime\prime}$($\pm$0.12)$\times$0.18$^{\prime\prime}$($\pm$0.08) & 0.85$^{\prime\prime}$$\times$0.64$^{\prime\prime}$\\
\enddata
\tablenotetext{a}{For the fittings, we use the data represented in Figure 1.}
\end{deluxetable}


\clearpage

\begin{deluxetable}{crrrr}
\tabletypesize{\scriptsize}
\tablecaption{Column Densities\tablenotemark{a}\label{tbl-1}}
\tablewidth{0pt}
\tablehead{
\colhead{Species} &\colhead{$N$} & \colhead{$X_{CH_3OH}$} \\
 & \footnotesize[cm$^{-2}$] & \\
}
\startdata
CH$_3$OH & 3.9$_{-1.8}^{+3.6}\times 10^{16}$ & --- \\
CH$_3$CH$_2$CN & 3.7$_{-0.5}^{+0.8}\times 10^{14}$ & 0.0094$_{-0.0051}^{+0.011}$ \\
HCOOCH$_3$ & 3.0$_{-0.6}^{+1.0}\times 10^{15}$ & 0.076$_{-0.044}^{+0.11}$\\
CH$_3$OCH$_3$ & 1.2$_{-0.4}^{+0.6}\times 10^{16}$ &0.31$_{-0.19}^{+0.54}$ \\
NH$_2$CHO & 1.1$_{-0.3}^{+0.4}\times 10^{14}$ &0.0027$_{-0.0016}^{+0.0041}$ \\
CH$_3$CHO & 3.4$_{-1.3}^{+2.4}\times 10^{14}$ &0.0087$_{-0.0059}^{+0.016}$\\
CH$_2$DCN & 2.3$_{-0.3}^{+0.5}\times 10^{13}$ &0.0006$_{-0.0003}^{+0.0007}$\\
D$_2$CO & 5.7$_{-2.5}^{+4.0}\times 10^{13}$ &0.0015$_{-0.0010}^{+0.0031}$ \\
\enddata
\tablenotetext{a}{The transition lines used for the derivation are the same as those represented in Figure 1.}
\end{deluxetable}

\clearpage

\begin{deluxetable}{crrr}
\tabletypesize{\scriptsize}
\tablecaption{Fitting results of partition functions\label{tbl-1}}
\tablewidth{0pt}
\tablehead{
\colhead{Species} &\colhead{$a$} & \colhead{$b$} & \colhead{Reference\tablenotemark{a}}\\
}
\startdata
CH$_3$OH & 0.3463 & 1.784 & 1\\
CH$_3$CH$_2$CN & 7.12725 & 1.5007 & 1 \\
HCOOCH$_3$ & 9.3414 & 1.748 & 2 \\
CH$_3$OCH$_3$ & 6.4669 & 2.1021 & 1 \\
NH$_2$CHO & 3.1792 & 1.4121 & 1 \\
CH$_3$CHO & 3.6211 & 1.7648 & 2 \\
CH$_2$DCN & 2.8865 & 1.4114 & 1 \\
D$_2$CO & 1.389 & 1.3844 & 1 \\
DNC & 1.7712 & 1.0061 & 1 \\
\enddata
\tablenotetext{a}{1:CDMS(http://www.astro.uni-koeln.de/cdms), 2:JPL(http://spec.jpl.nasa.gov/home.html)}
\end{deluxetable}

\clearpage





\begin{thebibliography}{}

\bibitem[Aikawa et al.(2008)]{aik08} Aikawa, Y., Wakelam, V., Garrod, R. T., Herbst, E. 2008, \apj, 674, 984
\bibitem[Aikawa et al.(2012)]{aik12} Aikawa, Y., Wakelam, V., Hersant, F., et al. 2012, \apj, 760, id. 40
\bibitem[Bacmann et al.(2003)]{Bac03} Bacmann, A., Lefloch, B., Ceccarelli, C., Steinacker, J., et al. 2003, \apj, 585, L55
\bibitem[Belloche et al.(2016)]{Bel16} Belloche, A., M\"uller, H. S. P., Garrod, R. T., Menten, K. M. 2016, \aap, 587, id. A91
\bibitem[Beuther et al.(2007)]{Beu07} Beuther, H., Zhang, Q., Bergin, E. A., et al. 2007, \aap, 468, 1045
\bibitem[Bisschop et al.(2007)]{Bis07} Bisschop, S. E., Jorgensen, J. K., van Dishoeck, E. F., de Wachter, E. B. M. 2007, \aap, 465, 913
\bibitem[Blake et al.(1996)]{Bla96} Blake, G. A., Mundy, L. G., Carlstrom, J. E., et al. 1996, \apjl, 472, L49
\bibitem[Bonnell \& Bate(2006)]{2006MNRAS.370..488B} Bonnell, I.~A., \& Bate, M.~R.\ 2006, \mnras, 370, 488 
\bibitem[Caselli et al.(1993)]{cas93} Caselli, P., Hasegawa, T. I., Herbst, E. 1993, \apj, 408, 548
\bibitem[Churchwell et al.(1990)]{chu90} Churchwell, E., Walmsley, C. M., Cesaroni, R. 1990, \aaps, 83, 119
\bibitem[Condon et al.(1998)]{con98} Condon, J. J., Cotton, W. D., Greisen, E. W., et al. 1998, \aj, 115, 1693
\bibitem[Feng et al.(2015)]{fen15} Feng, S., Beuther, H., Henning, Th., et al. 2015, \aap, 581, id. A71
\bibitem[Foster et al.(2012)]{fos12} Foster, J. B., Stead, J. J., Benjamin, R. A., et al. 2012, \apj, 751, id. 157
\bibitem[Foster et al.(2014)]{fos14} Foster, J. B., Arce, H. G., Kassis, M., et al. 2014, \apj, 791, id. 108
\bibitem[Fuente et al.(2001)]{fue01} Fuente, A., Neri, R., Mart\'in-Pintado, J., et al. 2001, \aap, 366, 873
\bibitem[Fuente et al.(2014)]{fue14} Fuente, A., Cernicharo, J., Caselli, P., et al. 2014, \aap, 568, id. A65
\bibitem[Garay et al.(2004)]{gar04} Garay, G., Fa\'{u}ndez, S., Mardones, D., et al. 2004, \apj, 610, 313
\bibitem[Garrod and Herbst(2006)]{gar06} Garrod, R. T., Herbst, E. 2006, \aap, 457, 927
\bibitem[Garrod (2013)]{gar13} Garrod, R. T. 2013, \apj, 765, id. 60
\bibitem[Garrod (2017)]{gar17} Garrod, R. T., Belloche, A., Mueller, H. S. P., Menten, K. M. 2017, \aap, 601, id. A48
\bibitem[Gerin et al. (1992)]{ger92} Gerin, M., Combes, F., Wlodarczak, G., et al. 1992, \aap, 259, L35
\bibitem[Goldsmith and Langer (1999)]{gol99} Goldsmith, P. F., Langer, W. D. 1999, \apj, 517, 209
\bibitem[Graninger et al. (2014)]{gra14} Graninger, D. M., Herbst, E., Oberg, K. I., Vasyunin, A. I. 2014, \apj, 787, id. 74
\bibitem[Herbst and van Dishoeck(2009)]{her09} Herbst, E., van Dishoeck, E. F. 2009, \araa, 47, 427
\bibitem[Hidaka et al.(2009)]{hid09} Hidaka, H., Watanabe, M., Kouchi, A., Watanabe, N. 2009, \apj, 702, 291
\bibitem[Hirota et al.(1998)]{hir98} Hirota, T., Yamamoto, S., Mikami, H., Ohishi, M. 1998, \apj, 503, 717
\bibitem[Hirota et al.(2002)]{hir02} Hirota, T., Ito, T., Yamamoto, S. 2002, \apj, 565, 359
\bibitem[Jaber et al.(2014)]{jab14} Jaber, A. A., Ceccarelli, C., Kahane, C., Caux, E. 2014, \apj, 791, id. 29
\bibitem[Jaber et al.(2014)]{jab14} J{\o}rgensen, J. K., van der Wiel, M. H. D., Coutens, A., et al. 2016, \aap, 595, id. A117
\bibitem[Kong et al.(2016)]{kon16} Kong, S., Tan, J. C., Caselli, P., et al. 2016, \apj, 821, id. 94
\bibitem[Kurayama et al.(2011)]{kur11} Kurayama, T., Nakagawa, A., Sawada, S. S., et al. 2011, \pasj, 63, 513
\bibitem[Kuchar and Bania(1994)]{kuc94} Kuchar, T. A., Bania, T. M. 1994, \apj, 436, 117
\bibitem[Maret et al.(2004)]{mar04} Maret, S., Ceccarelli, C., Caux, E., et al. 2004, \aap, 416, 577
\bibitem[Mookerjea et al.(2007)]{moo07} Mookerjea, B., Casper, E., Mundy, L. G., Looney, L. W. 2007, \apj, 659, 447
\bibitem[Nummelin et al.(1998)]{Num98} Nummelin, A., Bergman, P., Hjalmarson, Å., et al. 1998, \apjs, 117, 427
\bibitem[Ohashi et al.(2016)]{2016ApJ...833..209O} Ohashi, S., Sanhueza, P., Chen, H.-R.~V., et al.\ 2016, \apj, 833, 209 
\bibitem[Oya et al.(2016)]{oya16} Oya, Y., Sakai, N., L\'{o}pez-Sepulcre, A., et al. 2016, \apj, 824, 88
\bibitem[Palau et al.(2011)]{pal11} Palau, A., Fuente, A., Girart, J. M., et al. 2011, \apjl, 743, L32
\bibitem[Parise et al.(2006)]{par06} Parise, B., Ceccarelli, C., Tielens, A. G. G. M., et al. 2006, \aap, 453, 949
\bibitem[Persson et al.(2018)]{per18} Persson, M. V., J{\o}rgensen, J. K., M\"{u}ller, H. S. P., et al. 2018, \aap, 610, id. A54
\bibitem[Qin et al.(2010)]{qin10} Qin, S.-L., Wu, Y., Huang, M., et al. 2010, \apj, 711, 399
\bibitem[Qin et al.(2015)]{qin15} Qin, S.-L., Schilke, P., Wu, J., et al. 2015, \apj, 456, 2681
\bibitem[Qin et al.(2016)]{qin16} Qin, S.-L., Schilke, P., Wu, J., et al. 2016, \apj, 803, id. 39
\bibitem[Rodgers and Charnley(2001)]{rod01} Rodgers, S. D., Charnley, S. B. 2001, \apj, 546, 324
\bibitem[Rodgers and Millar(1996)]{rod96} Rodgers, S. D., Millar, T. J. 1996, \mnras, 280, 1046
\bibitem[Rathborne et al.(2005)]{rat05} Rathborne, J. M., Jackson, J. M., Chambers, E. T., et al. 2005, \apjl, 630, L181
\bibitem[Rathborne et al.(2006)]{rat06} Rathborne, J. M., Jackson, J. M., Simon, R. 2006, \apj, 641, 389
\bibitem[Sanhueza et al.(2010)]{san10} Sanhueza, P., Garay, G., Bronfman, L., et al. 2010, \apj, 715, 18
\bibitem[Sanhueza et al. (2012)]{san12} Sanhueza, P., Jackson, J. M., Foster, J. B., et al. 2012, \apj, 756, id. 60
\bibitem[Sanhueza et al.(2017)]{2017ApJ...841...97S} Sanhueza, P., Jackson, J. M., Zhang, Q., et al.\ 2017, \apj, 841, 97 
\bibitem[Sakai et al. (2008)]{sak08} Sakai, T., Sakai, N., Kamegai, K., et al. 2008, \apj, 678, 1049
\bibitem[Sakai et al. (2013)]{sak13} Sakai, T., Sakai, N., Foster, J. B., et al. 2013, \apjl, 775, id. L31
\bibitem[Sakai et al. (2015)]{sak13} Sakai, T., Sakai, N., Furuya, K., et al. 2015, \apj, 803, id. 70
\bibitem[Sakai and Yamamoto (203)]{nsak13} Sakai, N., Yamamoto, S. 2013, Chemical Reviews, 113, 8981
\bibitem[Sutton et al. (1985)]{sut85} Sutton, E. C., Blake, G. A., Masson, C. R., Phillips, T. G. 1985, \apjs, 58, 341
\bibitem[Schilke et al. (1992)]{sch92} Schilke, P., Walmsley, C. M., Pineau Des Forets, G., et al. 1992, \aap, 256, 595
\bibitem[Tafalla and Santiago(2004)]{taf04} Tafalla, M., Santiago, J. 2004, \aap, 414, L53
\bibitem[Taquet et al. (2012)]{Taq15} Taquet, V., Ceccarelli, C., Kahane, C. B. 2012, \apj, 748, id. L3
\bibitem[Taquet et al. (2015)]{Taq15} Taquet, V., L\'opez-Sepulcre, A., Ceccarelli, C., et al. 2015, \apj, 804, id. 81
\bibitem[Turner (1991)]{tur91} Turner, B. E. 1991, \apjs, 76, 617
\bibitem[van der Tak et al. (2007)]{van07} van der Tak, F. F. S., Black, J. H., Sch\"{o}ier, F. L., et al. 2007, \aap, 468, 627
\bibitem[Viti and Williams(1999)]{vit99} Viti, S., Williams, D. A. 1999, \mnras, 305, 755
\bibitem[Wang et al.(2010)]{2010ApJ...709...27W} Wang, P., Li, Z.-Y., Abel, T., \& Nakamura, F.\ 2010, \apj, 709, 27
\bibitem[Xu et al. (2016)]{xu16} Xu, J.-L., Li, D., Zhang, C.-P., et al. 2016, \apj, 819, id. 117
\bibitem[Yanagida et al.(2014)]{yan14} Yanagida, T., Sakai, T., Hirota T., et al. 2014, \apjl, 794, id. L10
\bibitem[Zhang et al.(2009)]{2009ApJ...696..268Z} Zhang, Q., Wang, Y., Pillai, T., \& Rathborne, J.\ 2009, \apj, 696, 268
\bibitem[Zhang et al.(2015)]{2015ApJ...804..141Z} Zhang, Q., Wang, K., Lu, X., \& Jim{\'e}nez-Serra, I.\ 2015, \apj, 804, 141 

\end{thebibliography}
\end{document}